\documentclass[bibyear]{aa} 

\usepackage{natbib,twoopt}
\usepackage[breaklinks=true]{hyperref} 
\bibpunct{(}{)}{;}{a}{}{,}             
\makeatletter
  \newcommandtwoopt{\citeads}[3][][]{\href{http://adsabs.harvard.edu/abs/#3}%
    {\def\hyper@linkstart##1##2{}%
     \let\hyper@linkend\@empty\citealp[#1][#2]{#3}}}
  \newcommandtwoopt{\citepads}[3][][]{\href{http://adsabs.harvard.edu/abs/#3}%
    {\def\hyper@linkstart##1##2{}%
     \let\hyper@linkend\@empty\citep[#1][#2]{#3}}}
  \newcommandtwoopt{\citetads}[3][][]{\href{http://adsabs.harvard.edu/abs/#3}%
    {\def\hyper@linkstart##1##2{}%
     \let\hyper@linkend\@empty\citet[#1][#2]{#3}}}
  \newcommandtwoopt{\citeyearads}[3][][]%
    {\href{http://adsabs.harvard.edu/abs/#3}
    {\def\hyper@linkstart##1##2{}%
     \let\hyper@linkend\@empty\citeyear[#1][#2]{#3}}}
\makeatother

\begin{document}

   \title{High-resolution He  \textsc{I}  10830 \AA\ narrowband imaging for precursors of chromospheric jets and their \textbf{quasi-periodic} properties}

   \author{Ya Wang\inst{1}; Qingmin Zhang\inst{1}; Zhenxiang Hong\inst{1}; Jinhua Shen\inst{2}; Haisheng Ji\inst{1}; Wenda Cao\inst{3}}
 
   \institute{\inst{1}Key Laboratory of Dark Matter and Space Astronomy, Purple Mountain Observatory, CAS, Nanjing, 210023, People's Republic of China \\
    \inst{2}Xinjiang Astronomical Observatory, CAS, Urumqi, 830011, People's Republic of China \\
     \inst{3}Big Bear Solar Observatory, New Jersey Institue of Technology, Big Bear City, CA 92314, USA  \\
                  \email{wangya@pmo.ac.cn}}

   \date{Received; accepted}
   \titlerunning{High-resolution He  \textsc{I}  10830 \AA\ narrowband imaging for the precursor of chromospheric jets and the quasi periodic property}
   \authorrunning{Y. Wang; Q. Zhang; Z. Hong, et al. }

\abstract{Solar jets are well-collimated plasma ejections in the solar atmosphere. They are prevalent in active regions, the quiet Sun, and even coronal holes. They display a range of temperatures, yet the nature of the cool components has not been fully investigated. In this paper, we show the existence of the precursors and quasi-periodic properties for two chromospheric jets, mainly utilizing the He \textsc{i} 10830 \AA\ narrowband filtergrams taken by the Goode Solar Telescope (GST). The extreme ultraviolet (EUV) counterparts present during the eruption correspond to a blowout jet (jet 1) and a standard jet (jet 2), as observed by  the Atmospheric Imaging Assembly (AIA) on board the Solar Dynamic Observatory (SDO). The high-resolution He \textsc{i} 10830 \AA\ observation captures a long-lasting precursor for jet 1, signified by a series of cool ejections. They are recurrent jet-like features with a quasi-period of about five minutes. On the other hand, the cool components of jet 2, recurrently accompanied by EUV emissions, present a quasi-periodic behavior with a period of about five minutes. Both the EUV brightening and He \textsc{i} 10830 \AA\ absorption show that there was a precursor for jet 2 that occurred about five minutes before its onset. We propose that the precursor of jet 1 may be the consequence of chromospheric shock waves, since the five-minute oscillation from the photosphere can leak into the chromosphere and develop into shocks. Then, we find that the quasi-periodic behavior of the cool components of jet 2 may be related to magnetic reconnections modulated by the oscillation in the photosphere. 
} 

   \titlerunning{High-resolution He  \textsc{I}  10830 \AA\ narrowband imaging for chromospheric jets }
   \maketitle

\section{Introduction} \label{intro}
 
Jets are common in the solar atmosphere, occurring from the photosphere to the corona and featuring a range of temperature components  \citep{2016SSRv..201....1R}. They are widely distributed in active regions, quiet regions \citep[e.g.,][]{2016ApJ...832L...7P, 2020ApJ...894..104P, 2021ApJ...918L..20H}, and polar regions \citep[e.g.,][]{2008ApJ...680L..73P, 2008ApJ...682L.137R, 2009ApJ...691...61P}. They are beams of plasma moving along an open magnetic field, usually associated with microflares, radio type III bursts, and so on. Jets could also occur in closed-field structures, such as loops \citep{2008A&A...491..279C}. Frequent jet activities are considered to be associated with coronal heating, the occurrence of coronal mass ejections (CMEs), filament oscillations, magnetohydrodynamic waves, solar wind, and other phenomena \citep[e.g.,][]{2014Sci...346A.315T, 2018ApJ...869...39M, 2019Sci...366..890S}. \\

Jets were  observed early as H$\alpha$ surges from ground-based telescopes \citep{1973SoPh...28...95R}. With space instruments, they have been observed in X-ray and EUV passbands, thus, they have been dubbed X-ray jets  \citep{1992PASJ...44L.173S} and EUV jets  \citep[e.g.,][]{1999SoPh..190..167A,2007A&A...469..331J,2009SoPh..259...87N,2011A&A...531L..13I}. With improved resolution, many small-scale jet-like events have been reported, such as spicules and dynamic fibrils, possibly formed in the similar mechanism as those of large-scale jets \citep{2021RSPSA.47700217S}. Recent high-resolution observations have revealed numerous transient, small-scale, collimated outflows, denoted as jetlets \citep{2014ApJ...787..118R}, at the base of coronal plume. These so-called jetlets also have an important role in heating the upper atmosphere \citep[e.g.,][]{2014ApJ...787..118R,2018ApJ...868L..27P,2021ApJ...907....1U,2022ApJ...933...21K}. \\

Coronal jets can be classified as standard jets and blowout jets \citep{2010ApJ...720..757M}. These definitions have been given in terms of the morphological description of coronal jets observed by Solar-B pre-launch/X-ray Telescope (Hinode/XRT) \citep{2016SSRv..201....1R}. Standard jets are typical anemone jets, with $\lambda$ types or inverted-Y structures in morphology. They are generally interpreted as a small-scale magnetic bipole reconnecting with the ambient open magnetic field \citep{1992PASJ...44L.173S}. The standard jet model can be used to explain X-ray and EUV jets based on the atmospheric layer where the magnetic reconnection had occurred \citep{2007Sci...318.1591S}. By comparison, blowout jets have mainly five characteristics \citep{2010ApJ...720..757M,2013ApJ...769..134M}: (1) the core of magnetic arcades is highly sheared, highly deviating from the potential field; (2) an extra jet's spire; (3) magnetic lines that participate in the eruption process of the jet being in front of the magnetic arcades; (4) additional brightenings inside the arch base of jets besides the outside brightenings; and (5) usually accompanied by the eruption of a twisting mini-filament \citep{2021RSPSA.47700217S}. \\

Solar jets are closely related to the evolution of local magnetic structures. Observations show that jets tend to occur above small opposite-polarity magnetic elements and regions of moving magnetic features \citep[e.g.,][]{1973SoPh...28...95R}. In addition, a large number of jets are caused by mini-filament eruptions or micro-eruptions that may be triggered by magnetic reconnection, accompanied by magnetic field cancellation \citep[e.g.,][]{2012ApJ...745..164S,2016ApJ...832L...7P,2018ApJ...864...68S}. High-resolution observations that include magnetic field data are needed to resolve the fine processes for different kinds of jets.\\

Flares, filament eruptions, and CMEs usually have precursors, such as transient small brightenings and the appearance of the so-called sigmoid magnetic structure \citep[e.g.,][]{
1985SoPh...97..387H, 
2007A&A...472..967C, 
2008A&A...481L..65M, 
2017ApJ...835...43S, 
2018ApJ...859..148W}. To understand the mechanism of jets, observations of their early signatures are essential, such as satellite sunspots $\footnote{The satellite sunspot is defined as polarity reversals near the edges of large-spot penumbrae .\citep[e.g.,][]{1968IAUS...35...77R}}$
 \citep[e.g.,][]{1968IAUS...35...77R, 
2015ApJ...815...71C, 2017PASJ...69...80S, 
2019ApJ...877...61M}, mini-filaments \citep[e.g.,][]{
2014ApJ...796...73H, 
2016ApJ...830...60H, 
2016ApJ...828L...9S, 
2017ApJ...835...35H}, coronal bright points \citep[e.g.,][]{
1977ApJ...218..286M, 2015ApJ...814..124C}, and micro-sigmoids \citep[e.g.,][]{2010ApJ...718..981R, 2000ITPS...28.1786C}. Some studies show that the jets associated with satellite sunspots are often closely related to the converging and shearing motions of small opposite-polarity magnetic elements \citep{2012ApJ...745..164S,2016ApJ...832L...7P,2018ApJ...853..189P}. The blowout jets are thought to be driven by a mini-filament eruption, similar to those observed in large-scale filament or CME eruptions \citep{2000ApJ...530.1071W}. In addition, solar jets are also observed being ejected from coronal bright points and small-scale sigmoids, which resemble the large-scale coronal sigmoid being the progenitor of solar eruptions. However, all the early signatures mentioned above are signals from other features, rather than directly from the jets themselves. In this regard, high-resolution observations are strongly needed. As we will see in this paper, high-resolution observations at He \textsc{i} 10830 \AA\ is capable of tracing the precursors of jet eruptions.\\
 
Solar jets tend to recur in the same place \citep[e.g.,][]{1999ApJ...513L..75C, 2007A&A...469..331J, 2008A&A...491..279C, 
2014A&A...567A..11Z}. Some of the recurrent jets exhibit a quasi-periodic behavior with periods ranging from several minutes to dozens of minutes. \citep{2007A&A...469..331J,2015Ap&SS.359...44L}. \citet{1999ApJ...513L..75C} analyzed several jets repeatedly occurred in the active region where a pre-existing magnetic flux was canceled by newly emerging flux with opposite flux. \citet{2008A&A...491..279C} found a correlation between recurring magnetic cancellation and X-ray jet emission. The recurrent jet emission was caused by chromospheric evaporation flows due to recurring magnetic reconnection. Recurrent jets can also be produced by some moving magnetic features \citep{2007ApJ...656.1197B}. In a previous paper, we discovered a periodic behavior of about five minutes for a small-scale jet, observed in the He \textsc{i} 10830 \AA\ passband \citep{2021ApJ...913...59W}. In addition, some micro-eruptions observed in the same passband also present a five-minute quasi-periodic behavior \citep{2022ApJ...928..153H}.  High-resolution observations at He \textsc{i} 10830 \AA\ have  rarely been used in studying solar jets. The results suggest that previously so-called recurrent EUV jets are related to the photospheric global oscillation if we consider their cool components to be better revealed by high-resolution imaging at 10830 \AA.\\


A subsequent question considers whether the five-minute oscillation generally exists in all chromospheric jets and whether it would depend on the type of jet. Furthermore, we asked how the oscillation in the photosphere might affect the chromospheric jets. Before these fundamental questions can be addressed, further studies are certainly needed. With the well-precessed high-resolution He \textsc{i} 10830 \AA\ observations from GST, we selected two chromospheric jets so that we could study their precursors and quasi-periodic properties, and to find similarities among them. In Section 2, we briefly introduce the data and instruments. We present our observation and data analysis in Section 3. Our conclusions and discussions are presented in Section 4. \\

\section{Data and instruments} \label{sec:style}

The data set includes narrowband filtergrams (bandpass: 0.5 \AA) at He \textsc{i} 10830 \AA\ and broadband filtergrams (bandpass: 10 \AA) at TiO 7057 \AA\ obtained from GST \citep{2010AN....331..620G} at Big Bear Solar Observatory (BBSO) on July 22, 2011, observed by \citet{2012ApJ...750L..25J}. The He \textsc{i} 10830 \AA\ data have a pixel size of $\sim$ 0$\farcs$09, a cadence of $\sim$ 14 s, and a field of view (FOV) of 90$\arcsec$$\times$90$\arcsec$. While the TiO 7057 \AA\ data have a pixel size of $\sim$0$\farcs$036 and a cadence of $\sim$15 s. TiO images provide the photospheric features in the region of jets and can be used for coalignment.\\

The EUV/UV images were observed by the AIA \citep{2012SoPh..275...17L} on board the SDO, with a pixel size of 0.$\arcsec$6. AIA covers six EUV and two UV passbands, with a cadence of 12 s and 24 s, respectively. The magnetograms are obtained from the Helioseismic and Magnetic Imager \citep[HMI; ][]{2012SoPh..275..229S} on board the SDO. The hmi.M$_{-}$45s data series provided us with full-disk magnetograms along the line of sight with a cadence of 45 s. 
 The photospheric dopplergrams with a cadence of 45 seconds were provided by hmi.V$_{-}$45s data series.\\ 

The full-disk H$\alpha$ images are obtained from the Global Oscillation Network Group \citep[GONG,][]{1994SoPh..152..321H}, with a pixel size of 1$\arcsec$ and a cadence of about 1 minute. GONG has the ability to obtain nearly continuous observations of the Sun on the ground and includes six stations: Big Bear Solar Observatory, High Altitude Observatory, Learmonth Solar Observatory, Udaipur Solar Observatory, Instituto de Astrof\'{i}sica de Canarias, and Cerro Tololo Interamerican Observatory. We made the alignment through the same features of the moss region based on the H$\alpha$ and He \textsc{ii} 304 \AA\ images from SDO/AIA.\\

To study small-scale jet events, a precise coalignment among the images of the photosphere, chromosphere, and corona is required. The He \textsc{i} 10830 \AA\ line is optically thin in most quiet regions, which allows the photosphere to be visible as a background. This is a benefit for the coalignment between TiO and He \textsc{i} 10830 \AA\ images. Meanwhile, the coalignment between TiO 7057 \AA\ broadband images and SDO/HMI continuum images was performed by using the same characteristics from sunspots, pores, and bright granules. The remaining offset after coalignment is less than $\sim$0$\farcs$5 \citep{2017RAA....17...25H}.\\

\section{Observation and data analysis} 

The two jets occurred in the NOAA active region 11259. Fig. 1 gives an overview of the region of interest with a field of view (FOV) of 60$\arcsec$$\times$60$\arcsec$. The red boxes  in panels (e1) and (e2) indicate the regions where the two jets occurred. White boxes roughly delineate the base boundaries of the two jets. The sunspot is dominated by a negative magnetic field, with some scattered positive polarity fields around it (panels d1 and d2). On the eastern side of the sunspot, there lies a EUV moss region (panel e2), where many periodic small-scale mass ejections of about five minutes, powered by magnetoacoustic oscillations, have been reported \citep[e.g.,][]{2021RAA....21..179J,2021RAA....21..105H}. The two jets are located on the northern side of the sunspot.\\

\begin{figure}
\centering
\includegraphics[width=9cm]{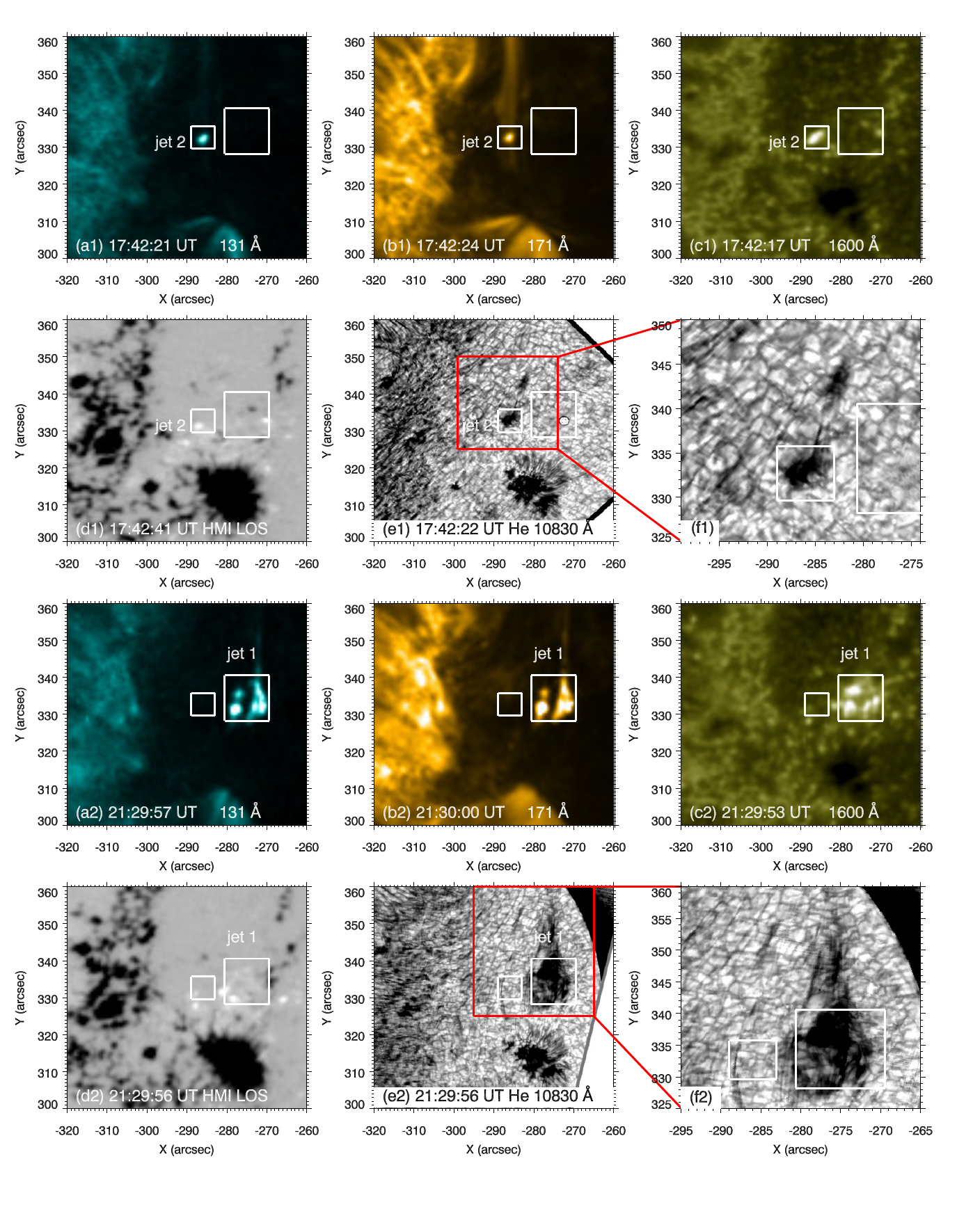}
\caption{Snapshots of the two jets (Panels a1-c1 and a2-c2) as observed by SDO/AIA at 131, 171 and 1600 \AA\ passbands. Panels (d1) and (d2) present corresponding magnetograms observed by HMI. Panels (e1) and (e2) are chromospheric images observed from BBSO/GST at He \textsc{i} 10830 \AA. Panels (f1) and (f2) display enlarged images for the two jets with the same FOV as the red box. White boxes give the areas for obtaining time profiles.} 
\label{fig1}
\end{figure}

\subsection{Jet1: Blowout jet and its precursor} 
GST carried out the observation for jet 1 from 20:47:45 UT to 21:34:35 UT, on July 22, 2011. Fig. 2 (a) gives the He \textsc{i} 10830 \AA\ image for jet 1. The EUV images in 131, 171, and 304 \AA\ passbands are presented in Fig. 2 (c-e). The blowout jet has multiple hot ejections and consists of two strands as observed in EUV images. Their speeds are estimated as 62-77 km s$^{-1}$ and 122-128 km s$^{-1}$, for strands 1 and 2, respectively. The appearance of the jet, as observed in high-resolution He \textsc{i} 10830 \AA\ images, consists of many thin threads that overall form the anemone-like morphology. Based on the EUV light curves given in Fig. 2 (L1), we find that there are two main eruptions about 10 minutes apart. EUV observations show that the jet started at $\sim$21:17:00 UT, and peaked at 21:21:36 and 21:31:20 UT, respectively. \\

Remarkably, the chromospheric counterpart of the blowout jet possesses a rich precursor stage lasting at least 30 minutes as shown on the time profile of He \textsc{i} 10830 \AA\ in Fig. 2 (L2) from $\sim$20:47 UT to $\sim$21:17 UT. The integrated intensity of He \textsc{i} 10830 \AA\ has already started to descrease at 20:47 UT. The enhanced absorption indicates the overpopulation of triplet helium. As shown in Fig. 3 (a1-a5), the precursor presents recurrent jet-like structures.  For comparison, we observed only a mass of dark materials on the H$\alpha$ images obtained from GONG in panels (b1-b5). Except for a brightening at 1600 \AA\ and emission at 304 \AA\ at $\sim$21:09:29 UT,  we found no other signals in all the EUV passbands from SDO/AIA during the precursor stage. \\

The corresponding animation displays the precursor stage clearly. The jet was preceded by slight disturbances observed on the integrated intensity curves of He \textsc{i} 10830 \AA. There are five intermittent jet-like features (labeled as ``1''-``5'' in Fig. 2 (L1)) shown as enhanced absorptions. Each enhanced absorption corresponds to intensified ejections of cool materials around the whole jet region. The corresponding picture is given in Fig. 3 (a1-a5). We performed a Morlet wavelet analysis \citep{1998BAMS...79...61T} for the detrended He \textsc{i} 10830 \AA\ light curve to determine the period of oscillation (Fig. 4). The wavelet power spectra (Fig. 4c) show the dominant period of about five minutes that is above the 95\% significance level.\\

The evolution and eruption of the blowout jet are shown in Figs. 5 and 6. The cool component is observed as a surge in H$\alpha$ images. For this blowout jet, the cool component is formed in the middle of the two strands, which refer to the hot components of the jet. For comparison, the surge consists of two cool ejections observed at He \textsc{ii} 304 \AA. The two cool ejections are also observed at He \textsc{i} 10830 \AA, and one ejection experienced a lateral motion (Fig. 6 (a5)). We thus consider how these two cool ejections were formed. The high-resolution observation in He \textsc{i} 10830 \AA\ passband shows that it is the result of an ejection of cool components from the lower atmosphere.\\

In the EUV passbands, this ejection is manifested as two strands at two different locations (Fig. 6), in agreement with the blowout jet model. However, in the He \textsc{i} 10830 \AA\ counterpart, the jet behaves differently. The two-part structure is contained in one anemone-shaped structure with enhanced absorption. Until $\sim$21:34:20 UT, one ejection has a lateral expansion. At $\sim$21:28 UT, we observe small dark arcades emerging inside the anemone structure (Fig. 6 (a1)). The emerging was accompanied by brightenings in 1600 \AA\ passband at the same location. The emerging arcades reconnect with the surrounding open magnetic field. The reconnection created a darkened cusp-shaped structure (Fig. 6 (a2)), which was expelled as a cool component along with the hot ejections in EUV images (Fig. 6 (c3-c5)), giving rise to strand 1. We find obvious material motions around the structure before the hot eruption. We believe that the magnetic field of the thin threads with enhanced absorption on the eastern side (pointed by red arrows) plays a part in the magnetic reconnection.  During the eruption, we observe brightenings in 1600 \AA\ passband (Fig. 6 (b2-b3)) inside and outside the base of the jet, and even He \textsc{i} 10830 \AA\ brightens at the footpoint. The ejection of the cool component looks like an eruption of a mini-filament in EUV counterparts. However, we see no filament before the eruption of the blowout jet in 304 \AA\ images. The cool materials may come from the lower atmosphere brought up by the emerging magnetic arcades.\\

In addition, we found that strand 2 in the EUV images displays multiple ejections. It has a rotational motion and a lateral expansion at 21:32:20 and 21:34:20 UT, respectively, which can be seen in panels (c4-c5) of Fig. 6 and the corresponding animation. Following the lateral expansion of hot ejections, an extra cool ejection appeared in He \textsc{i} 10830 \AA\ at 21:34:22 UT, as indicated by a white arrow in Fig. 6 (a5). In this evolutionary process, the enhanced absorption turns into emission, as observed as brightenings in front of the arch base of the jet (seen in Fig.6 (a1-a5)). Strand 2 was ejected faster, with a speed reaching more than 120 km s$^{-1}$.\\

\begin{figure}
\centering
\includegraphics[width=9cm]{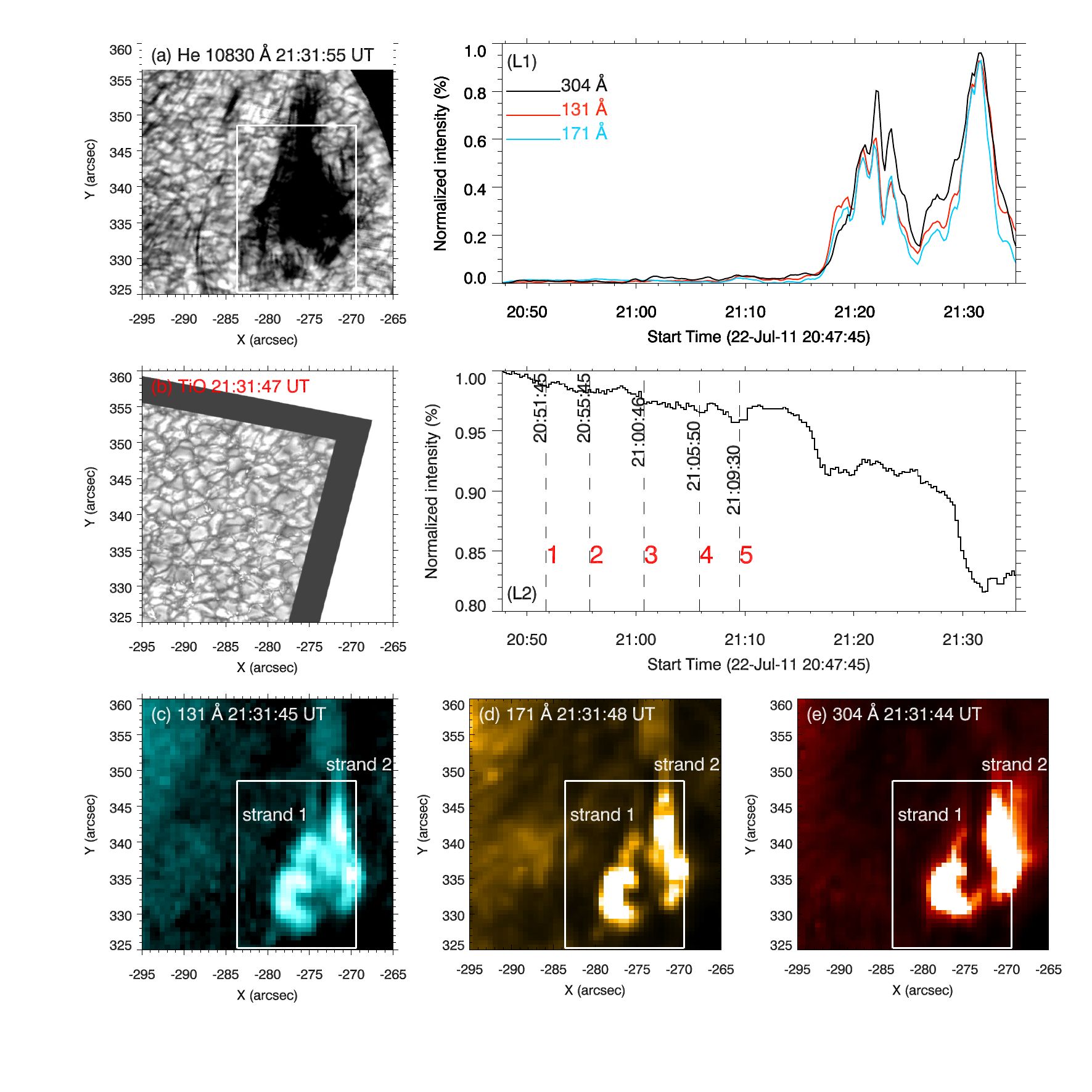}
\caption{Morphology of jet 1 as observed in He \textsc{i} 10830, TiO 7057, 131, 171, and 304 \AA\ passbands (Panels a-e). Time profiles for the integrated intensities of the jet in corresponding passbands are represented in panels (L1) and (L2). The integrated region for the time profiles is the jet region inside the white boxes in panels (a-e). An online animation of the figure is available; it lasts 32 seconds from 20:47:45 UT to 21:30:11 UT.} 
\label{fig2}
\end{figure}

\begin{figure}
\centering
\includegraphics[width=9cm]{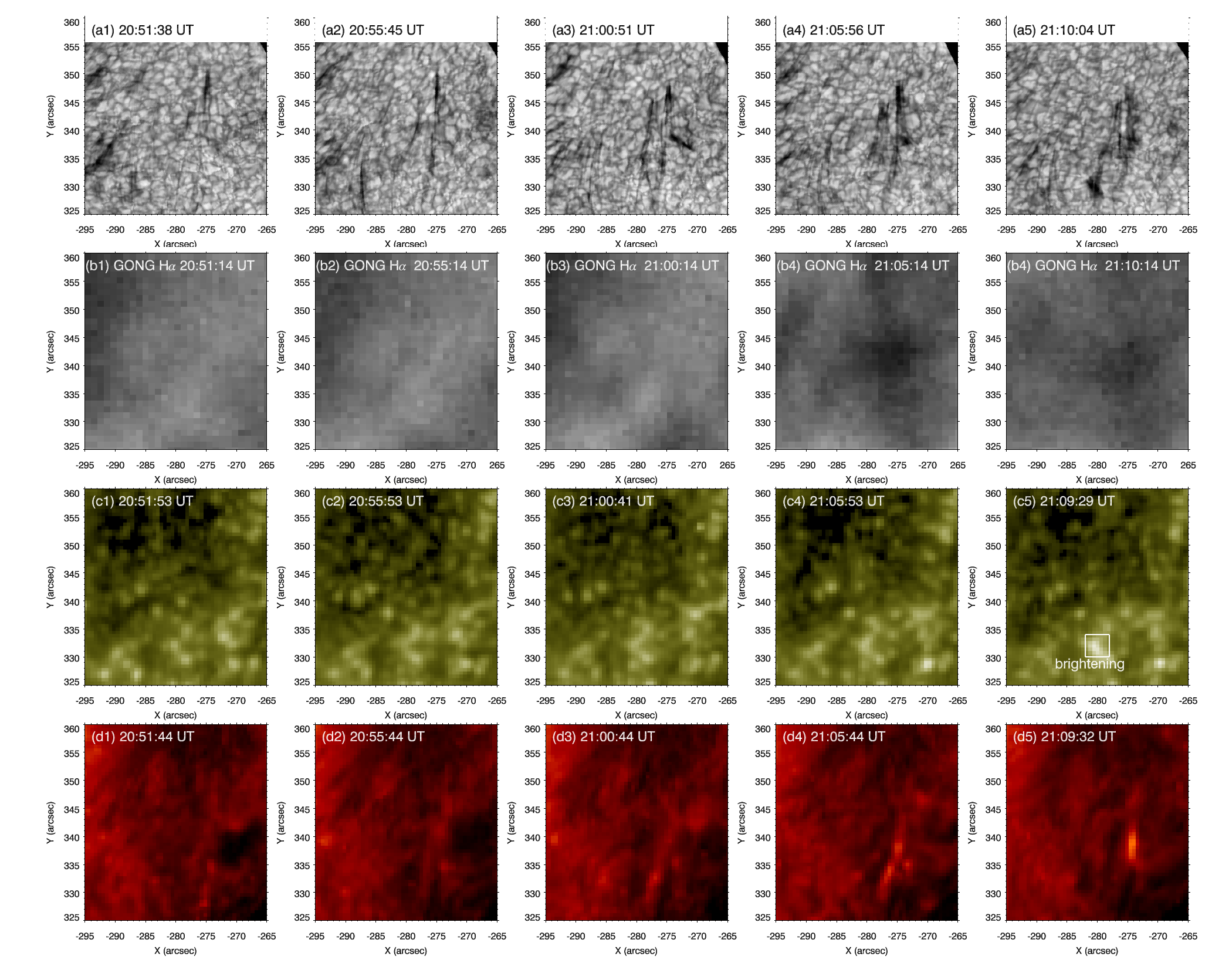}
\caption{Snapshots of the precursor phase of the blowout jet as observed in He \textsc{i} 10830, H$\alpha$ 6563, 1600, and He \textsc{ii} 304 \AA\ passbands from $\sim$20:51 UT to $\sim$21:10 UT. The times are chosen as the ones given by the vertical lines in Fig. 2 (L2).} 
\label{fig 3}
\end{figure}

\begin{figure}
\centering
\includegraphics[width=9cm]{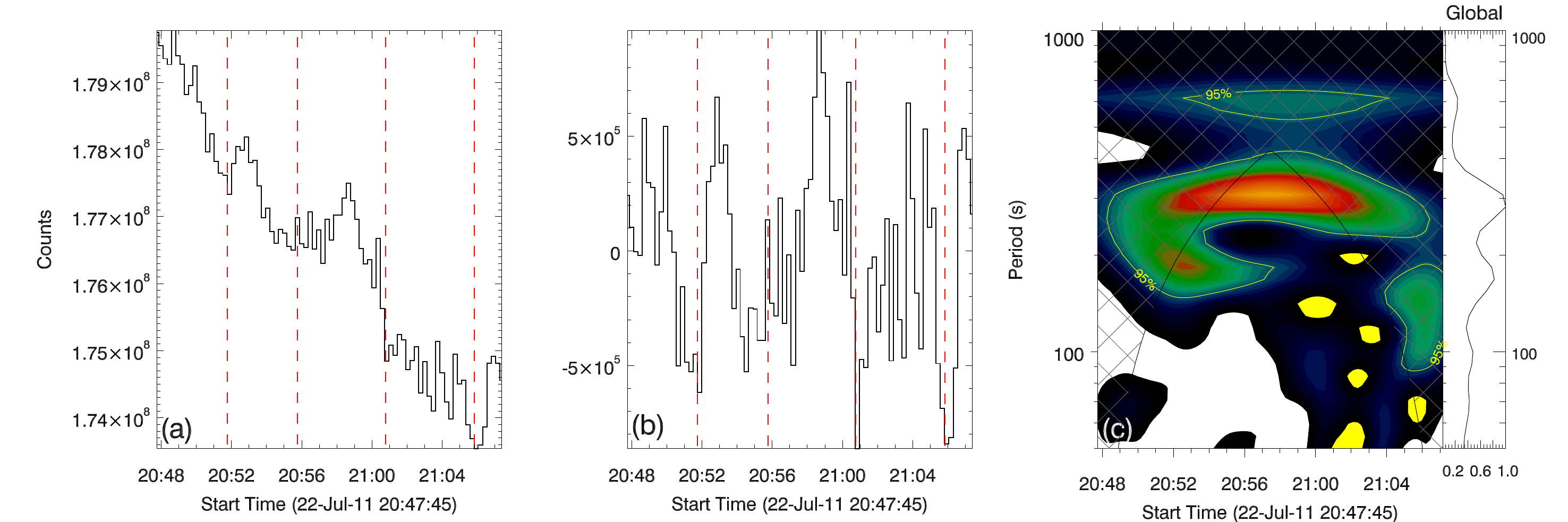}
\caption{Quasi-periodicity of the jet-like structure in the precursor phase of jet 1. Panel (a): Intensity variations in a selected region of the He \textsc{i} 10830 \AA\ image.  Red dashed lines indicate the enhanced absorptions at 20:51:45, 20:55:45, 21:00:46, and 21:05:50 UT. Panel (b): Detrended light curve after subtracting the smoothed light curve from the original intensity overlaid with the same set of dashed lines. Panel (c): Wavelet power spectrum of the detrended signal and global wavelet power spectrum. Contour outlines the 95\% significance level. } 
\label{fig 4}
\end{figure}

\begin{figure}
\centering
\includegraphics[width=9cm]{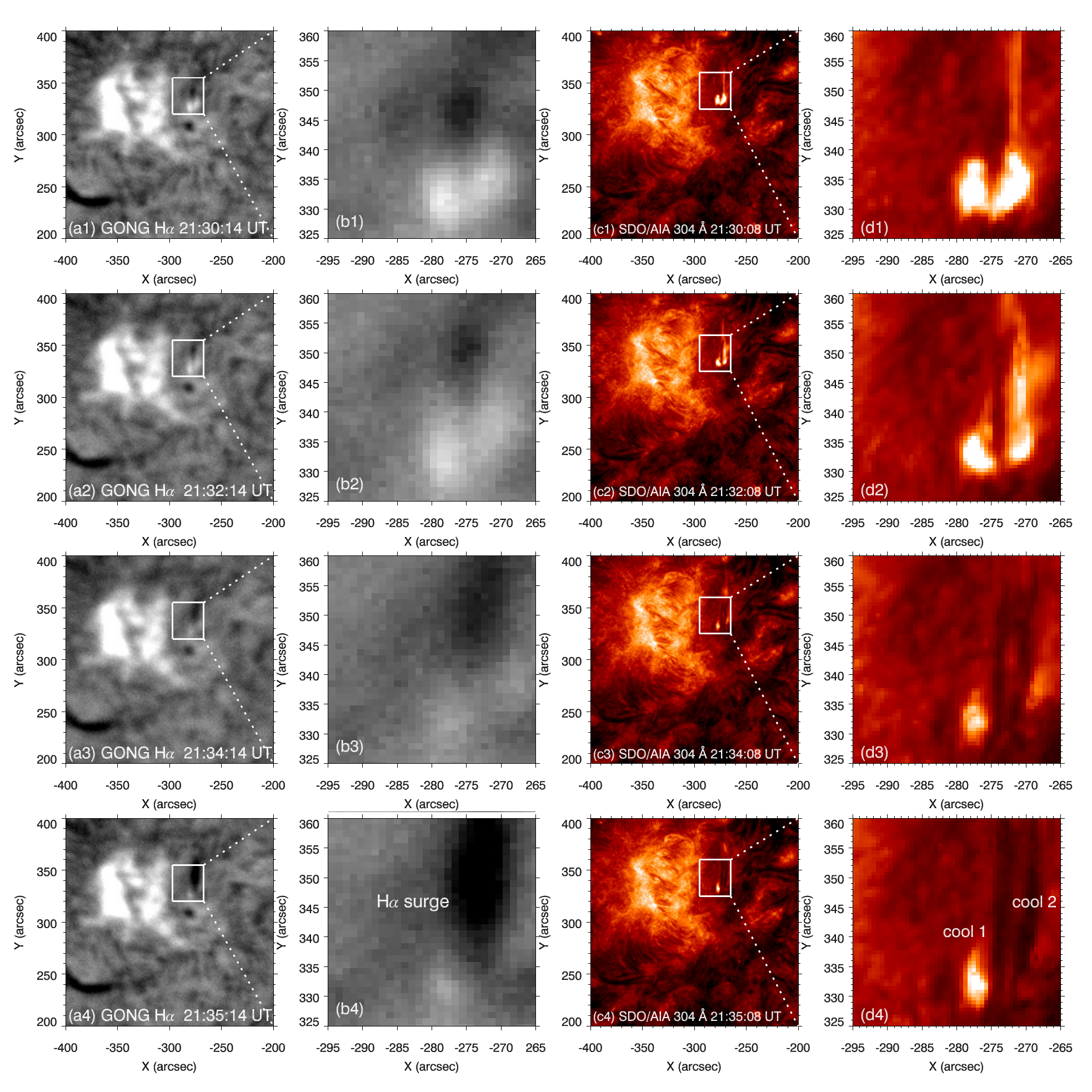}
\caption{Time sequence images of the jet region as observed in H$\alpha$ and He 304 \AA. Panels (b1-b4) and (d1-d4) are the zoomed-in images of the jet in the region of white boxes.} 
\label{fig 5}
\end{figure}

\begin{figure}
\centering
\includegraphics[width=9cm]{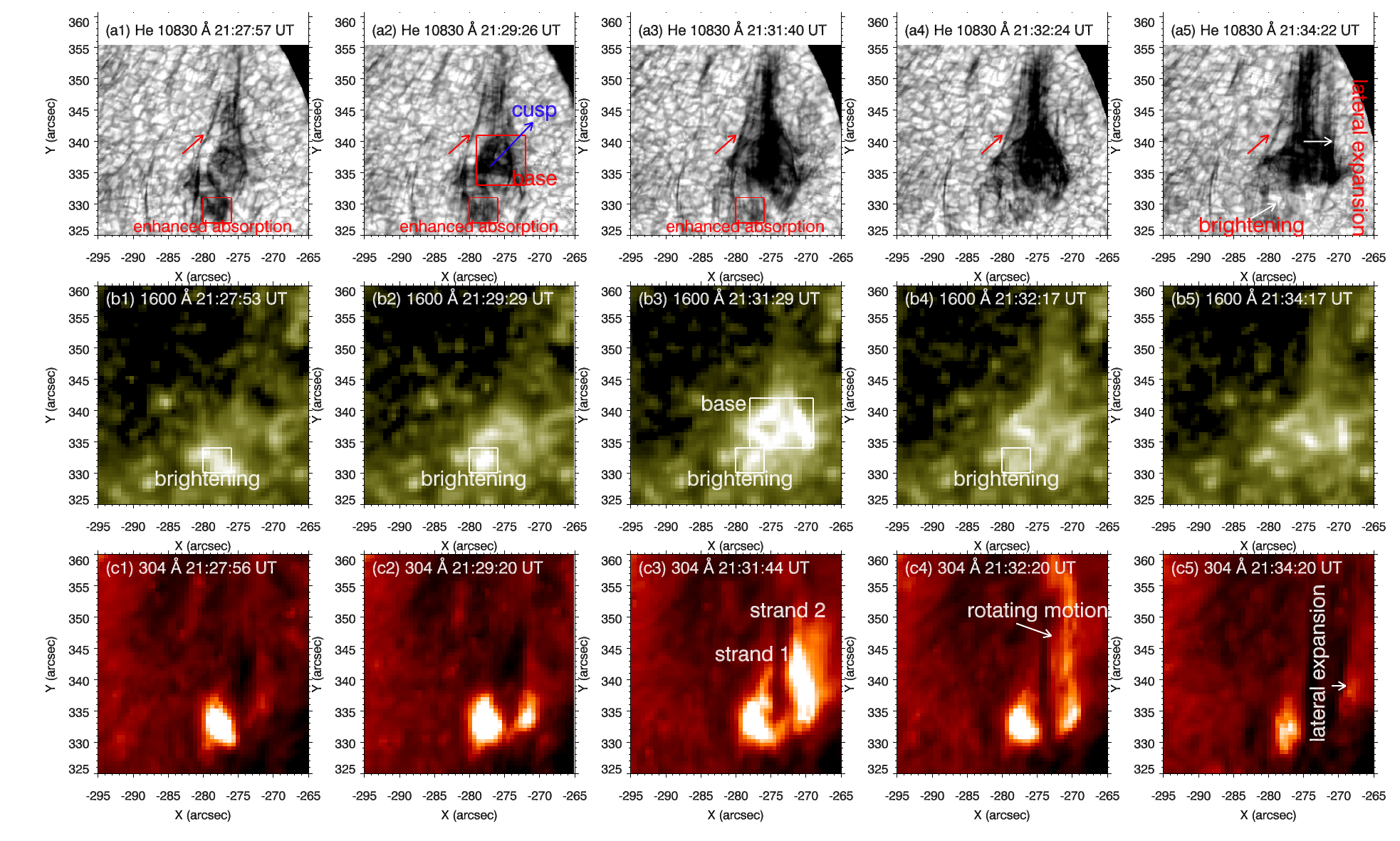}
\caption{Evolution of the \textbf{blowout} jet (jet 1) as observed in He \textsc{i} 10830, 1600, and 304 \AA\ passbands. } 
\label{fig 6}
\end{figure}

\subsection{Jet2: Standard jet and precursor} 

The second jet (jet 2) is shown in Fig. 7. The jet manifests itself as a EUV bright point with a size of about 2$\arcsec$$\times$2$\arcsec$ in AIA images. The jet was caused by magnetic reconnection between an emerging granule-sized magnetic element and the ambient magnetic field as studied by \citet{2013ApJ...769L..33Z}. Their time profiles of EUV passbands are presented in Fig. 7 (L1). We find that it is quasi-periodic and peaks at 17:42:00 UT. As shown in Fig. 7, the chromospheric jet observed at He \textsc{i} 10830 \AA\ displays enhanced absorption. At the peak time, the morphology of the jet in He \textsc{i} 10830 \AA\ images shows a domelike base with a distinct thin spire. It contains many thin and dark threads, and the width of a separated thread could be about 100 km as reported by \citet{2021ApJ...913...59W}. Along the ejection labeled by slice 1 in panel (a), we made a time-distance diagram and obtained the speed of the ejection, measured as 119 km s$^{-1}$. \\

For the light curves of integrated intensities shown in Fig. 7 (L1-L2), the He \textsc{i} 10830 \AA\ absorption valleys correspond to the emission peaks at 131, 171, and 304 \AA\ passbands. Before the jet was initiated, the jet-like structure in He \textsc{i} 10830 \AA\ images appeared as a precursor, as shown in Fig. 8 (a1). It corresponds to a valley on the lightcurve of He \textsc{i} 10830 \AA\ passband at $\sim$17:36 UT in Fig. 7 (L2). The EUV counterpart is relatively weak, yet there appeared as a EUV bright point. This phenomenon is similar to the jet-like structure that occurred in the precursor phase of jet 1.  However, the precursor phase of jet 1 lasted for half an hour and it appeared only once for jet 2. It is worth noting that the precursor proceeds about five minutes before the eruption of jet 2.\\

The moments of enhanced absorption of the jet or emission of the integrated intensity from the bright point, judging from the absorption valleys or emission peaks, appear at 17:36:00, 17:42:00, 17:45:30, and 17:49:10 UT, which are labeled with numbers from ``1'' to ``4.''  We give snapshots at the four moments above at He \textsc{i} 10830 \AA, H$\alpha$, He \textsc{ii} 304 \AA, and 94 \AA,\ as shown in Fig. 8. With the four enhancements of the cool materials, the corresponding EUV bright point was observed, indicating that it was heated to millions of degrees. According to the standard jet model, the material ejection is caused by magnetic reconnection between an emerging arcade and the ambient magnetic field. We also performed a Morlet wavelet analysis of the He \textsc{i} 10830 \AA\ light curve to determine the period of oscillation of jet 2 (Fig. 9a). The wavelet power spectrum (Fig. 9c) shows that the period of about five minutes is above the 95\% significance level.\\

To  detect their quasi-periodic properties, the time profiles and wavelet power spectrums of the photospheric doppler velocity (when picking out the redshift one that is positive) over the jet region were obtained using the SDO/HMI dopplergrams with a cadence of 45 seconds. Both of the wavelet power spectrums (Fig. 10) during the precursor of jet 1 and the whole process of jet 2 show a period of about five minutes from $\sim$20:47:56 UT to $\sim$21:07 UT and 17:35:11 UT to $\sim$18:00 UT, respectively. In addition, we compared the phase differences between the photospheric doppler velocity  and chromospheric jets, with the result shown in Fig. 10 (c1 and c2). The phase differences are different for the two jets. For the precursor of jet 1, the photospheric doppler redshift velocity is $\sim$107 seconds ahead of the jet oscillation. However, for jet 2, the photospheric doppler velocity lagged behind the jet oscillation by $\sim$119 seconds. We note that valleys of the chromospheric oscillation observed at He \textsc{i} 10830 \AA\ in Fig. 10 are the peak time of the jets in EUV passbands. On the other hand, the TiO images reflect the photospheric signal well. For the integrated intensity of TiO 7057 \AA\ in the jet region, it presents a quasi-periodic osillation with a period of five minutes as well. 


\begin{figure}
\centering
\includegraphics[width=9cm]{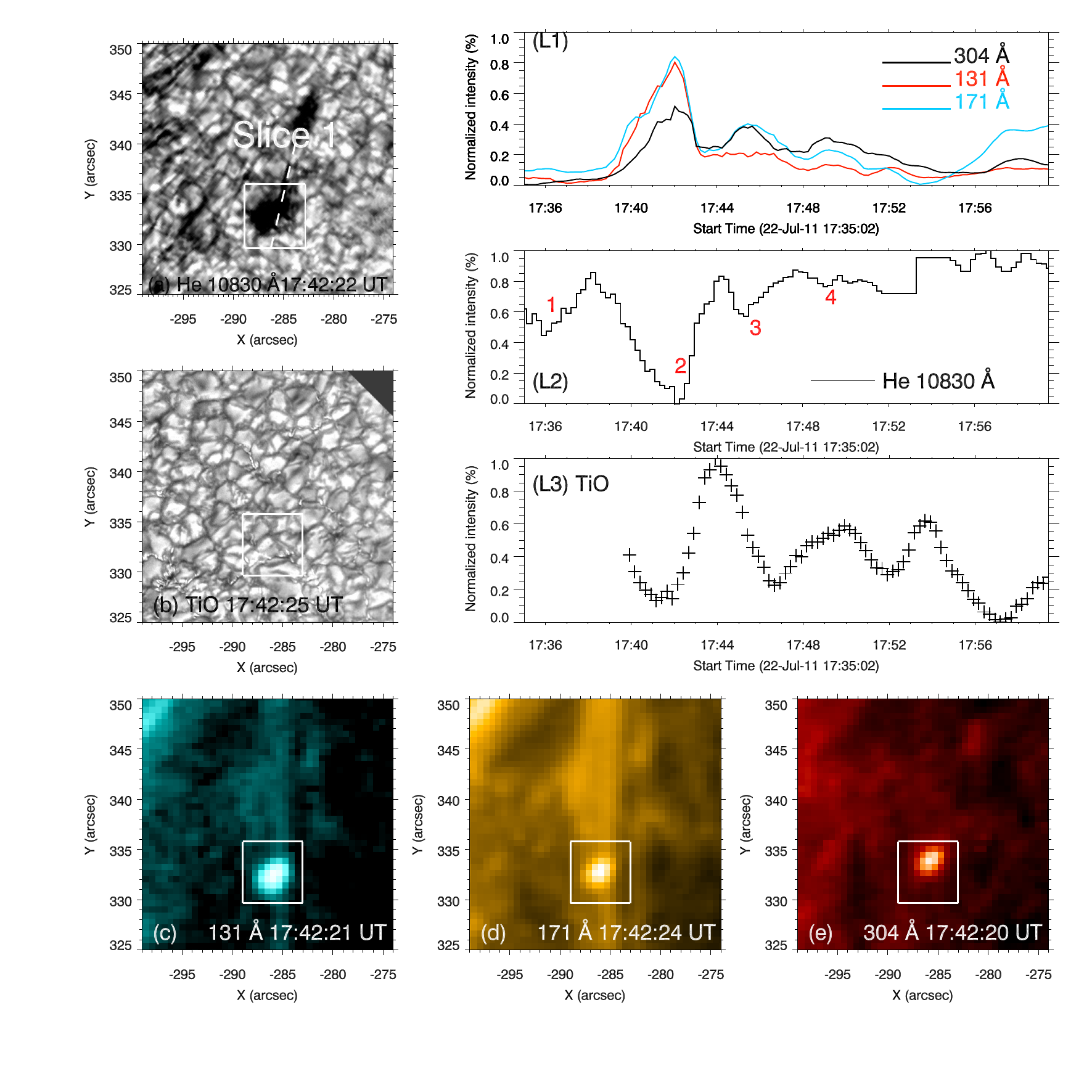}
\caption{Morphology and time profiles for jet 2. Panels a-e give the morphology of jet 2 as observed in He \textsc{i} 10830, TiO 7057, 131, 171, and 304 \AA\ passbands. The time profiles for the integrated intensities of the jet in corresponding passbands are represented in panels (L1), (L2), and (L3). The integrated region for the time profiles is the base of the jet inside the white boxes in panels a-e. An online animation of the figure is available; it lasts for 15 seconds from 17:35:00 UT to 17:55:56 UT.} 
\label{fig 7}
\end{figure}

\begin{figure}
\centering
\includegraphics[width=9cm]{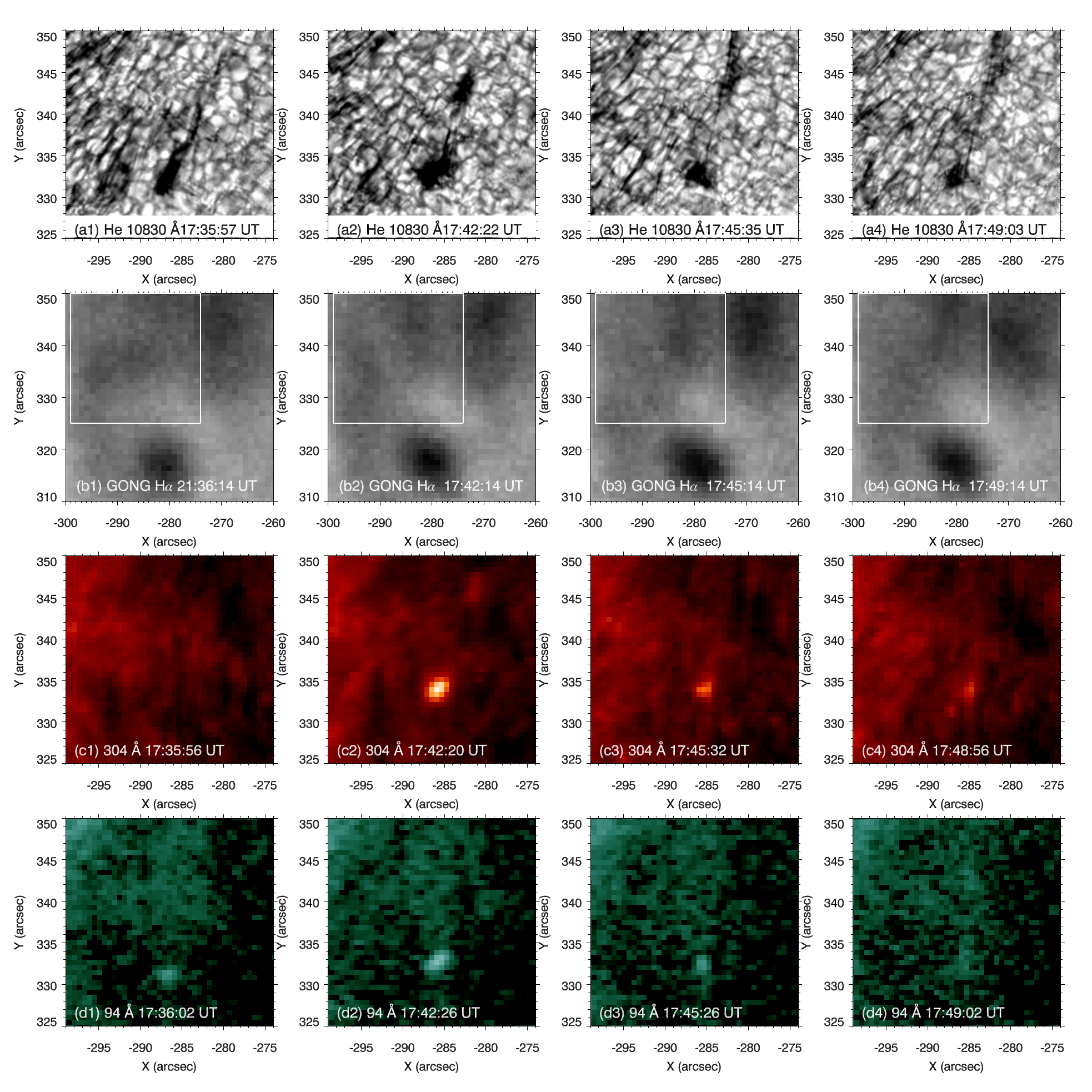}
\caption{Snapshots of the recurrent jet observed in He \textsc{i} 10830, H$\alpha$ 6563, He \textsc{ii} 304, 94 \AA\ passbands at four peak times 17:35:57, 17:42:22, 17:45:35, and 17:49:03 UT. White boxes in panels (b1-b4) indicate the FOV in the other panels.} 
\label{fig 8}
\end{figure}

\begin{figure}
\centering
\includegraphics[width=9cm]{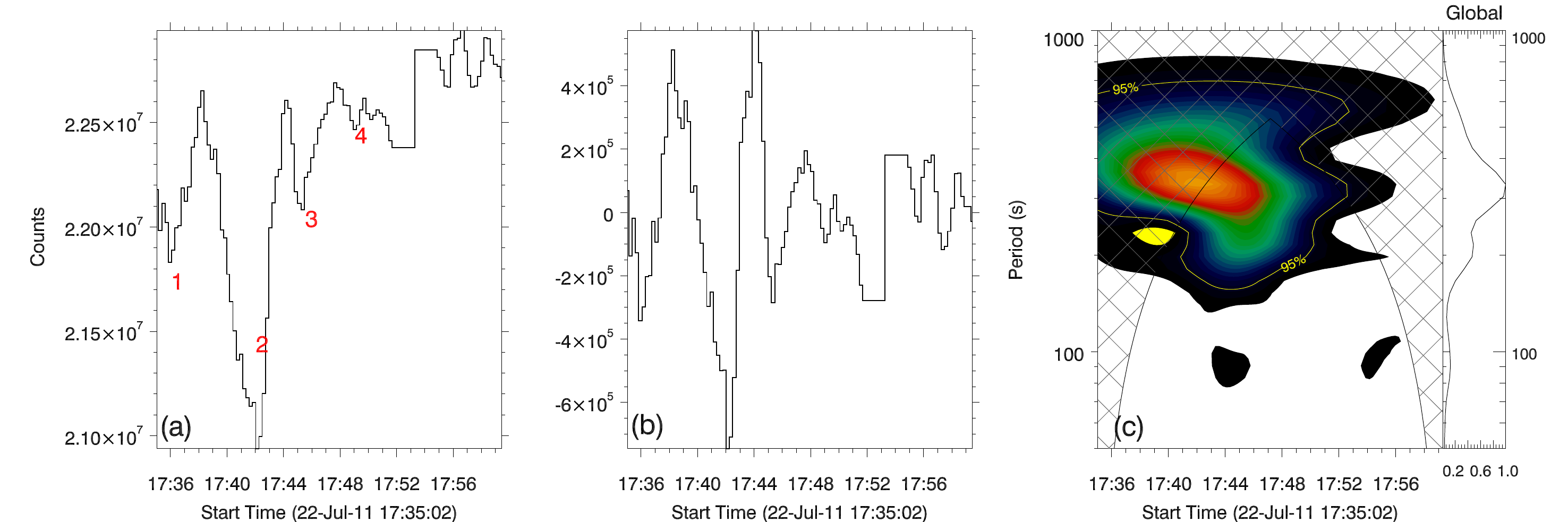}
\caption{Periodicity of jet 2. Panel (a): Intensity variations in a selected region of the He \textsc{i} 10830 \AA\ image.  Red numbers indicate the enhanced absorptions. Panel (b): Detrended light curve after subtracting the smoothed light curve from the original intensity. Panel (c): Wavelet power spectrum of the detrended signal and global wavelet power spectrum. Contour outlines the 95\% significance level.} 
\label{fig 9}
\end{figure}

\begin{figure}
\centering
\includegraphics[width=9cm]{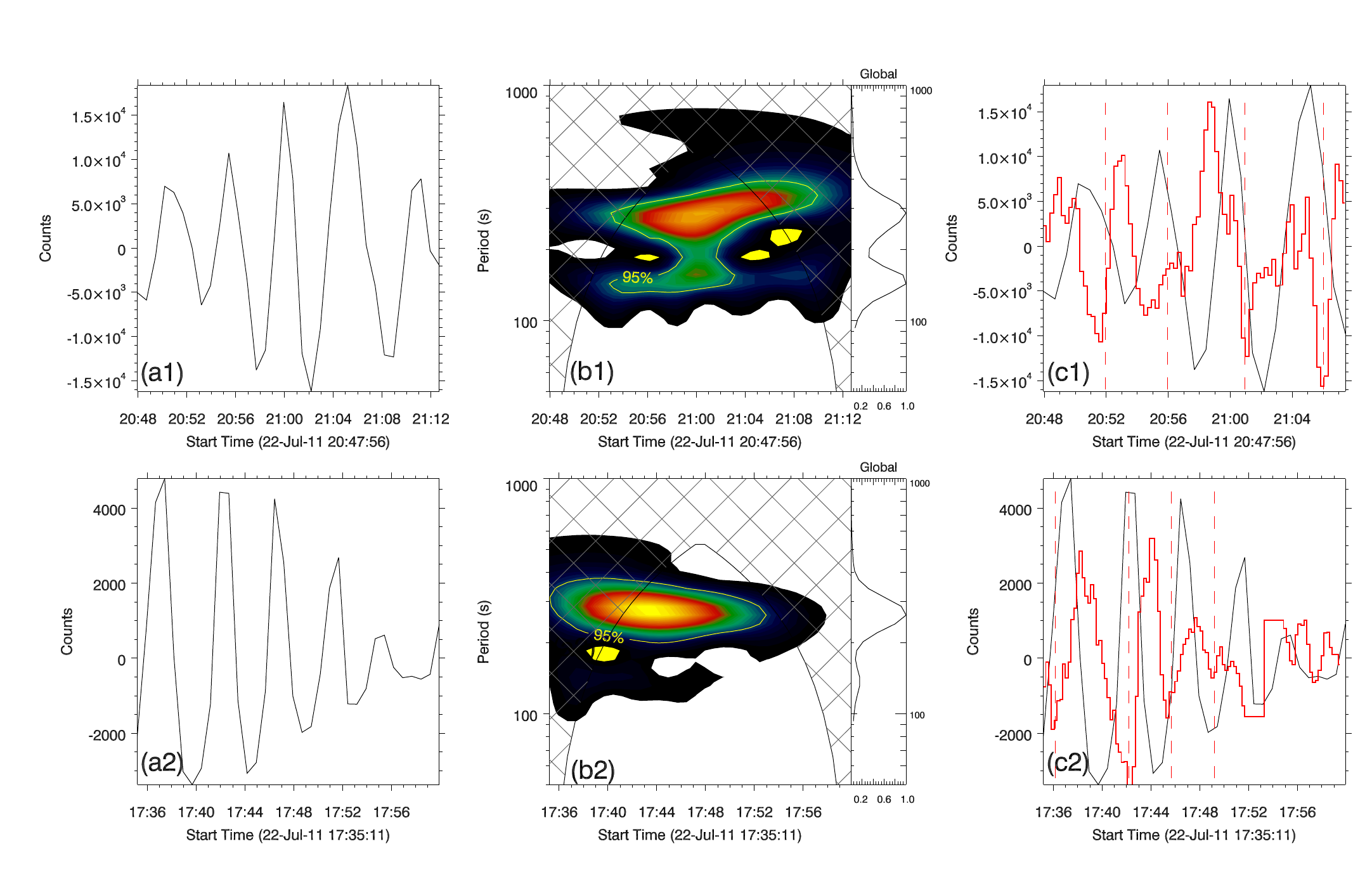}
\caption{Periodicity of the doppler redshift velocity from SDO/HMI in the regions of jet 1 (upper panels) and jet 2 (lower panels). Left panels: Detrended light curves after subtracting the smoothed light curves from the original light curves. Middle panels: Wavelet power spectrum and global wavelet power spectrum of the detrended light curves from the doppler redshift velocity. Contour outlines the 95\% significance level. Right panels:  Time profiles in left panels are superimposed with the detrended light curves of the two jets at He \textsc{i} 10830 \AA (red curves).} 
\label{fig 10}
\end{figure}




\section{Conclusions and discussions} 

In this paper, we analyzed two jets of two different types with the well-processed high-resolution He \textsc{i} 10830 \AA\ narrowband filtergrams taken by GST on July 22, 2011. Aided by EUV observations from AIA and photospheric velocity from HMI on board the SDO spacecraft, as well as the H$\alpha$ images from the GONG, we studied the precursors, fine structures, multi-temperature components, and the quasi-periodic oscillation of the two jets with two different types. \\

Our main conclusions for the two jets are as follows:
   \begin{enumerate}
      \item The chromospheric jet 1, of which the EUV counterpart is a blowout jet, possesses a precursor phase that lasted half an hour. The precursor presents jet-like dark and thin threads that come upward recurrently in the region of the blowout jet. The precursor is direct jet phenomenon rather than other features, such as satellite sunspots, mini filaments, or mini-sigmoids.  In addition, we find that the precursor is quasi-periodic, with a period of about five minutes. We confirm the five-minute oscillation with both wavelet analysis to its light curve and visible inspection to the recurrent appearance of dark threads in He \textsc{i} 10830 \AA\ images. During the precursor phase, there are few EUV emissions. Two strands in EUV images are observed during the eruption. One of them experiences a rotation motion and a lateral expansion. The cool materials observed as H$\alpha$ surge experience a lateral motion and split into two cool ejections, which are also observed in He \textsc{i} 10830 and He \textsc{ii} 304 \AA\ passbands.\\
      
      \item Jet 2 displays one enhanced absorption about five minutes before the eruption, which can also be regarded as the precursor of the jet. For its EUV counterpart, it presents a coronal bright point. We find a total of four enhanced absorptions in the available data range of He \textsc{i} 10830 \AA, which also shows a quasi-periodic property with a period of about five minutes. The five-minute oscillation is confirmed with both wavelet analysis and image inspection.\\
                 
\end{enumerate}
 Our key finding is the detection of quasi-periodical oscillations with a characteristic
period of about five minutes in the jet-like features that precede the jets.  \citet{2018ApJ...855L..21B} found that the vast majority of coronal jets (either single or recurrent) are preceded by slight precursor disturbances observed in the mean intensity, occurring several minutes before each intensive jet ejection. These authors considered these enhancements of brightness as precursors of coronal jets and detected quasi-periodical oscillations with characteristic periods from sub-minutes up to four minutes in the bright point brightness that precedes the jet eruption. \citet{2020A&A...639A..22J} studied the pre-jet phase of the jets and found quasi-periodic intensity oscillations at their base. The quasi-periodic intensity oscillations have a period of between two and six minutes, and are reminiscent of acoustic or magnetohydrodynamic waves. The value that both jets have in common is their quasi-periodic property, with a period of about five minutes. With regard the possible mechanism of these oscillations, we  discuss them below. \\

For the precursor of jet 1, it appears to be similar to the periodicities and oscillations in jet-like observations, as previously reported by several authors \citep[e.g.,][]{2004Natur.430..536D, 2006ApJ...647L..73H, 2006RSPTA.364..383D, 2007ApJ...655..624D}. 
 \citet{2004Natur.430..536D} and \citet{2006RSPTA.364..383D} reported that not all dynamic fibrils that belong to some kind of small-scale jet-like features on the disk show oscillations or recurrence, but a significant number of them do. Most of the power in dynamic fibrils possess periods between four and six minutes. \citet{2004A&A...423.1133T} also suggested a similar period range in mottles in the enhanced network or weak plages. They suggested that the formation of dynamic fibrils is the result of chromospheric shock waves driven by convective flows and global oscillations in the photosphere and the convection zone.  The mechanism was initially proposed by \citep{1964ApJ...140.1170P}. Later, the rebound shock model \citep{1982ApJ...257..345H, 
1989ApJ...343..985S, 1990ApJ...349..647S} suggests that upward shocks are observed as recurrent jets.  In addition, in a stratified medium, a sudden disturbance can make the medium oscillate at a natural frequency close to the acoustic cut-off frequency 
\citep{2015ApJ...808..118C}, which can nonlinearly develop into a sequence of upward shocks. These shocks may be observed as recurrent jets.  \\
      
 The oscillatory power spectrum of the photospheric driver is dominated by the power at five minutes. The chromosphere filters out waves with periods that are longer than the local acoustic cutoff period. However, the acoustic cutoff depends on the inclination of the magnetic field. In the inclined field, for example, at the edge of a plage region, the photospheric five-minute wave can leak into the chromosphere along the inclined magnetic line, developing into shocks and forming jet-like features. This scenario can be used to interpret the periodic jet-like structure with a period of five minutes in the precursor phase of the blowout jet.\\

      However,  the driving mechanism of the quasi-periodic property in solar jets is still a subject of debate. They may be different with respect to various magnetic topologies. For the standard jet (jet 2), the quasi-periodic oscillation occurred during the eruption of the jet and we observe brightenings in the EUV images simultaneously. These are the result of magnetic reconnection between magnetic arcades and ambient open magnetic field lines.  The quasi-periodic oscillation is more likely due to the magnetic reconnection process modulated by the five-minute oscillation in the photosphere. The quasi-periodic behavior indicates a modulation from p-mode oscillations to the reconnection, which has been modeled numerically by \citet{2006SoPh..238..313C}.\\
      
        The difference in the phase delay between the photospheric doppler velocity and the ejection of the two jets may indicate different physical mechanisms. For jet 1, the proposed mechanism is based on the finding that the five-minute oscillation can leak into the chromosphere, developing into shocks and forming jet-like features; whereas for jet 2, the quasi-periodic behavior may indicate a modulation from the photospheric oscillation to the magnetic reconnection.\\
      
     On the other hand, it is interesting to discuss the property of the cool material located in the middle of the two hot strands of the blowout jet. It has been widely reported that the eruption of mini-filament triggers jets and numerous observations also have confirmed the result. With regard to our result,  we consider whether these cool materials are a mini-filament or cool components of the jets -- or a mixture of them. We see no mini-filaments before the eruption of jet 1, at He 304 \AA\ or H$\alpha$ 6563 \AA. This may be due to the limitation of  the spatial resolution. In the He 10830 \AA\ passband, they present high-dynamic jet-like features with a periodicity of about five minutes. The formation of the cool ejection possibly shows up with the emerging magnetic arcades from below the photosphere. Thus, it is worth investigating what its appearance is in other wavelengths with high-resolution observations.\\
      
      In this paper, the analysis of five-minute oscillation of the chromospheric jets and their precursors is still limited to case studies. Further investigations of larger jet samples are needed to investigate whether a five-minute oscillation is a common phenomenon in solar jets when considering high-resolution observations from He \textsc{i} 10830 \AA\ imaging and to find out how the oscillation in the photosphere affects these jets. \\

\begin{acknowledgements}
We thank the anonymous referee for his/her constructive comments and suggestions. We thank the team of SDO/AIA and SDO/HMI for providing valuable data. The AIA and HMI data were downloaded via the Joint Science Operations Center. The H$\alpha$ data from GONG is downloaded via Virtual Solar Observatory. \\
This work is supported by the National Key R\&D program of China 2021YFA1600502 (2021YFA1600500). This work is supported by NSFC grants 12003072, 12173092, 12073081, 12273115, and 12273101. Y.W. is supported by the Youth Found of JiangSu No. BK20191108.  BBSO operation is supported by NJIT and US NSF AGS-1821294 grant. 
\end{acknowledgements}
 
\bibliographystyle{aaStyle}
\bibliography{refJet}

\end{document}